# Data-Sharing Relationships in the Web

Adriana Iamnitchi
[1]Computer Science Department
University of Chicago
Chicago, IL 60637

anda@cs.uchicago.edu

Matei Ripeanu
Computer Science Department
University of Chicago
Chicago, IL 60637

matei@cs.uchicago.edu

Ian Foster [1]
Mathematics and Computer Science
Argonne National Laboratory
Argonne, IL 60439

foster@mcs.anl.gov

## ABSTRACT
We propose a novel structure, the data-sharing graph, for characterizing sharing patterns in large-scale data distribution systems. We analyze this structure in two such systems and uncover small-world patterns for data-sharing relationships. Using the data-sharing graph for system characterization has potential both for basic science, because we can identify new structures emerging in real, dynamic networks; and for system design, because we can exploit these structures when designing data location and delivery mechanisms. We conjecture that similar patterns arise in other large-scale systems and that these patterns can be exploited for mechanism design.

## 1. INTRODUCTION
Studies show that the graph in which nodes are Web pages and edges are the associated hyperlinks has small-world properties [1, 2]. However, this static property is 'wired' in the Web structure and does not reflect usage patterns. Usage patterns are captured by the aggregate file popularity distribution which has been shown to follow a Zipf law for the Web [3]. However, this latter metric does not capture the mapping between users and the subset of files in which each of them is interested: the overall popularity of an individual file appears as an aggregate over all users in the system.

We study the relationships that form among users based on the data subsets in which they are interested. We capture and quantify these relationships by modeling the system as a *data-sharing graph*. To this end, we propose a new measure that captures common user interests in data and justify its utility with studies on two data-distribution systems: the Web and a high-energy physics collaboration. Our main finding is that small-world patterns form in the data-sharing graph.

We conjecture that similar patterns occur in other systems and that these patterns can be exploited to build efficient, decentralized data-location and data-delivery mechanisms.

## 2. THE DATA SHARING GRAPH
We define the data-sharing graph of a system as a graph whose nodes are the data consumers in that system, such as users or their machines' IP addresses. Edges connect pairs of nodes whose activity satisfies a similarity criterion, $C$: for example, they connect nodes that access at least $m$ common files during a time interval $T$. We shall later refine this definition on two concrete examples: the Web and a physics collaboration. We use these data-sharing graphs to identify data-sharing patterns and to evaluate how they vary with $C$ and $T$.

### 2.1 The Web Data Sharing Graph
For the study of Web traces, we consider a node as an IP address. An edge connects two nodes that fetched at least $p$ same pages or accessed at least $s$ common servers during a $T$ seconds interval. We vary $m$ from 1 to 500, $s$ from 1 to 10, and $T$ from 2 minutes to 4 hours.

We use the Boeing proxy traces [4] as a representative sample for the data access pattern in the Web. These traces represent a five-day record of all HTTP requests (more than 20M requests per day) from a large organization (Boeing) to the Web.

### 2.2 D0 Data Sharing Graph
The D0 Experiment [5] is a collaboration consisting of thousands of physicists at more than 70 institutions in 18 countries. These physicists mine a PetaByte (c.2003) of measured and simulated data. A typical job analyzes and produces new data files. We analyzed logs for the first six months of 2002, amounting to about 23,000 job runs submitted by more than 300 users and involving more than 2.5 millions requests for about 200,000 distinct files.

A node in the D0 data-sharing graph is a D0 user. An edge connects two users if they accessed at least *one* common file during a $T$-day interval. We vary $T$ from one day to a month.

### 2.3 Data Sharing Graph Characteristics
We discover that *both these data-sharing graphs display small-world properties*. Two characteristics differentiate small-world graphs when compared to random graphs of the same size: first, a small average path length, typical of random graphs; second, a significantly larger clustering coefficient that is independent of graph size. The clustering coefficient captures how many of a node's neighbors are connected to each other. One can picture a small world as a graph constructed by loosely connecting a set of almost complete subgraphs. Social networks, in which nodes are people and edges are relationships; the Web, in which nodes are pages and edges are hyperlinks; and neural networks, in which nodes are neurons and edges are synapses or gap junctions, are a few of the many examples of small-world networks [6].

The table presents the average path-length and the clustering coefficient (averaged over multiple intervals of equal length) of data-sharing graphs defined by a few different similarity criteria. We compare these metrics with those of random graphs of similar sizes. Note that despite diversity in systems, graph definitions (i.e., similarity criteria), and graph sizes, the values are remarkably close.

| Experiment and similarity criteria used | Graph size (avg.) | | Average path length (avg.) | | Clustering coefficient | |
|---|---|---|---|---|---|---|
| | # node | # links | DS graph | Rand. graph | DS graph | Rand graph |
| Web, $m$=1, $T$=2min. | 1542 | 38k | 2.89 | 2.61 | 0.782 | 0.033 |
| Web, $m$=10, $T$=30min | 5629 | 183k | 2.33 | 2.67 | 0.762 | 0.012 |
| Web, $m$=100, $T$=2h | 7856 | 178k | 2.35 | 3.26 | 0.753 | 0.004 |
| Web, $s$=10, $T$=5min | 1375 | 56k | 2.22 | 2.14 | 0.803 | 0.051 |
| D0, $m$=1, $T$=7 days. | 41 | 176 | 2.39 | 2.63 | 0.752 | 0.231 |

The figure compares these data-sharing graphs with a selection of well-known, small-world graphs, including citations network, power grid, movie actors, Internet, Web [7]. Axes represent ratios between the metrics of interest of these graphs and random graphs of the same size. As above, for our data sharing graphs, each point in the plot represents averages for all graphs

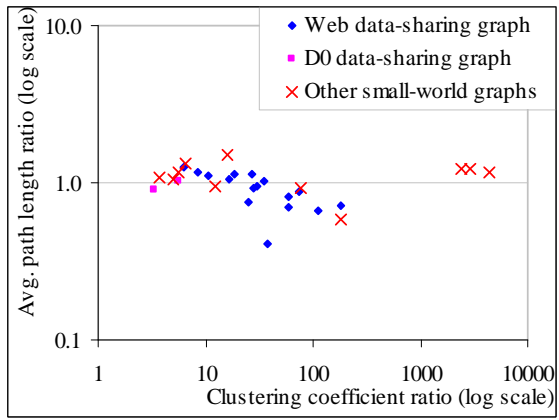

constructed from one similarity criterion.

We are not surprised that the data-sharing graphs resulting from the D0 experiment display small-world properties as they probably mirror collaboration relationships among scientists that have been documented, as many other social networks, as small world [8]. However, we do not yet have an explanation for the emergence of this property in the Boeing Web traces, other than that they also reflect the commonality of user interests. We shall study other traces to (in)validate this assumption.

## 3. SUMMARY AND SIGNIFICANCE
We study data-sharing patterns in large data-distribution systems. To this end we define the data-sharing graph and explore its characteristics. *This study is the first to reveal the small-world structure of data-sharing relationships among users*. We believe our results have implications for basic science (as we identify new structures emerging in real, dynamic networks) as well as for system design (as we can exploit these structures for mechanism design).

Two recent studies [9, 10] have focused on exploiting data-sharing patterns to improve the efficiency of file-location mechanisms. The same Boeing traces are used to drive their simulations. While these two studies intuit the existence of data-sharing sharing patterns, they do not identify or quantify them.

We observed small-world data-sharing patterns in two very different systems: the Web and a scientific collaboration. Our results lead us to conjecture that similar patterns exist in many other data-distribution systems. The challenge is now to learn how to exploit these patterns—for example, to (a) *build better location mechanisms,* as suggested in [9-11]; and/or (b) *build more efficient data delivery mechanisms*. Caching is generally employed in data distribution systems to save bandwidth and to reduce data access latency—for example, by placing proxy caches topologically 'close' to clients. In a data processing system, for example, where deriving new data implies significant computational effort, a group cache based not on proximity but on shared data usage could save CPU cycles and reduce latency in data delivery.